\documentclass[
a4paper, 12pt, ,nofootinbib, groupedaddress, superscriptaddress, amsmath, amssymb]{revtex4}
\usepackage{graphicx}
\usepackage{dcolumn}
\usepackage{bm}
\usepackage{amssymb}
\usepackage{amsmath}
\usepackage{epsfig}    
\usepackage{color}
\usepackage{slashed}
\usepackage{hhline}
\def\be{\begin{equation}}
\def\ee{\end{equation}}
\newcommand{\bea}{\begin{eqnarray}}
\newcommand{\eea}{\end{eqnarray}}
\newcommand{\nn}{\nonumber}

\numberwithin{equation}{section}

\begin{document}

{\begin{flushright}{KIAS-P16073}
\end{flushright}}

\title{A testable radiative neutrino mass model without additional symmetries
and Explanation for the $b \to s \ell^+ \ell^-$ anomaly}

\author{Kingman Cheung}
\email{cheung@phys.nthu.edu.tw}
\affiliation{Physics Division, National Center for Theoretical Sciences, 
Hsinchu, Taiwan 300}
\affiliation{Department of Physics, National Tsing Hua University, 
Hsinchu 300, Taiwan}
\affiliation{Division of Quantum Phases and Devices, School of Physics, 
Konkuk University, Seoul 143-701, Republic of Korea}

\author{Takaaki Nomura}
\email{nomura@kias.re.kr}
\affiliation{School of Physics, KIAS, Seoul 130-722, Korea}

\author{Hiroshi Okada}
\email{macokada3hiroshi@cts.nthu.edu.tw}
\affiliation{Physics Division, National Center for Theoretical Sciences, 
Hsinchu, Taiwan 300}

\pacs{}
\date{\today}

\begin{abstract}
We propose a one-loop radiative Majorana-type neutrino-mass matrix
without any kind of additional symmetries by introducing two
leptoquark-like bosons only. In this scenario, we show that the
anomaly appearing in the process $b\to s\ell\bar\ell$ can be explained
without any conflicts against various constraints such as
lepton-flavor violations, flavor-changing neutral currents, oblique
parameters $\Delta S,\ \Delta T$, and the Drell-Yan process. We make
the predictions for the flavor-violating lepton-pair production
($e\mu$, $e\tau$, and $\mu\tau$) at the LHC, as well as the cross
sections for pair production of these leptoquark-like bosons.
\end{abstract}
\maketitle

\section{Introduction}
The standard model (SM) of particle physics is so successful that all
the experiments searching for signs beyond the SM resulted in negative
results and the SM has been tested to a precision of $10^{-3}$.  Yet,
the neutrino oscillation experiments accumulated enough evidences that
the neutrinos do have masses.  Massive neutrinos are then the only
formally established evidences beyond the SM. Although there are some
other observations which also point to physics beyond the SM, such as
existence of dark matter, accelerated expansion of the Universe, and
matter-antimatter asymmetry, however, they are not as convincing
as the massive neutrinos. 

Extensions or modifications of the SM are often put forward to explain
the neutrino masses and their oscillation patterns. The most celebrated 
one is the see-saw mechanism with the introduction of heavy right-handed
(RH) neutrinos at the mass scale of $10^{11-12}$ GeV \cite{see-saw}.
There are variations in the see-saw type models with TeV RH neutrinos
\cite{tev-see-saw}. The advantage of TeV see-saw models is that they
can be tested in the LHC experiments \cite{lhc-see-saw}. 
Another type of neutrino mass models is based on loop diagrams, in which
the small neutrino mass is naturally obtained by the suppression loop factor.
Some classic examples are the one-loop Zee model \cite{zee} and 
Ma model \cite{ma}, two-loop Zee-Babu model~\cite{babu}, 
three-loop Krauss-Nasri-Trodden model \cite{knt}, etc.
Often in this type of models, some ad hoc symmetries are introduced to 
forbid some unwanted contributions or the see-saw contributions if there
are RH neutrinos in the model. 

Recently, there was an $2.6\sigma$ anomaly in lepton-universality violation
measured in the ratio 
$R_K \equiv B(B\to K\mu\mu)/B(B\to K ee) = 0.745 ^{+0.090}_{-0.074} \pm 0.036$ 
by LHCb \cite{lhcb-2014}. Also, sizable deviations from the SM prediction
were recorded in angular distributions of $B \to K^*\mu\mu$ \cite{lhcb-2013}.
The results can be accounted for by a large negative
contribution to the Wilson coefficient $C_9$ of the semileptonic operator
$O_9$, and also contributions to other Wilson coefficients, in particular to 
$C'_9$ {\cite{Descotes-Genon:2015uva, Hiller:2014yaa, Hiller:2016kry}}.

Here we propose a simple extensions of the SM 
with introduction of a
color-triplet $SU(2)_L$-doublet scalar boson $\eta$ and 
a color-antitriplet $SU(2)_L$-triplet scalar boson 
$\Delta$ without assuming
further discrete or gauge symmetries. The $SU(3)_C$, $SU(2)_L$, $U(1)_Y$ 
quantum numbers of the new fields are summarized in Table~\ref{tab:1}.
We shall show that this model can successfully explain the neutrino masses
and the oscillation pattern, as well as solving the anomalies
in $b \to s \ell \bar \ell$ with additional contributions to 
$C_{9,10}$ and $C'_{9,10}$, and at the same time
satisfying all the existing constraints of lepton-flavor violations (LFV),
flavor-changing neutral currents (FCNC), and $S,T,U$ parameters. 
Furthermore, the masses of $\eta$ and $\Delta$ bosons are in the
TeV scale, and so can be tested in the Drell-Yan process and
lepton-flavor violating production, and also directly in the pair 
production via the $\ell\ell jj$ final state.  This is the main 
result of the work.

This paper is organized as follows.
In Sec.~II, we review the model and describe the constraints.
In Sec.~III, we analyze numerically the parameter space so as to
solve the anomaly in $b\to s \ell\ell$, and calculate the cross sections
for lepton-flavor violating production.
We conclude in Sec.~IV.

\section{Model setup and Constraints}
\begin{table}[t!]
\begin{tabular}{|c||c|c|}
\hline\hline  
                   & ~$\eta$~  & ~$\Delta$ \\\hline 
$SU(3)_C$ & $\bm{3}$  & $\bar{\bm3}$  \\\hline 
$SU(2)_L$ & $\bm{2}$  & $\bm{3}$  \\\hline 
$U(1)_Y$   & $\frac16$ & $\frac13$    \\\hline
\end{tabular}
\caption{\small 
Charge assignments of the new fields $\eta$ and $\Delta$  
under $SU(3)_C\times SU(2)_L\times U(1)_Y$.}
\label{tab:1}
\end{table}

The new field contents and their charges are shown in Table~\ref{tab:1}, 
in which 
the color-triplet $\eta$ is an $SU(2)_L$ doublet with $1/6$ hypercharge, 
while the color-antitriplet $\Delta$ is an $SU(2)_L$ triplet with 
$1/3$ hypercharge.
The relevant Lagrangian for the interactions of the $\eta$ and $\Delta$ with
fermions and the Higgs field is given by 
\begin{align}
-\mathcal{L}_{Y}
&=
 f_{ij} \overline{d_{R_i}} \tilde \eta^\dag L_{L_j} + g_{ij} \overline{ Q^c_{L_i}}
 (i\sigma_2) \Delta L_{L_j} 
-\mu\Phi^\dag \Delta \eta+ {\rm h.c.},
\label{Eq:lag-flavor}
\end{align}
where $(i,j)=1-3$ are generation indices, 
$\tilde\eta\equiv i\sigma_2\eta^*$, $\sigma_2$ is the second 
Pauli matrix, and $\Phi$ is the SM Higgs field that develops a 
nonzero vacuum expectation value (VEV), which is symbolized by 
$\langle\Phi\rangle\equiv v/\sqrt2$. 
We work in the basis where all the coefficients are real and positive
for simplicity. 
The scalar fields can be parameterized as 
\begin{align}
&\Phi =\left[
\begin{array}{c}
w^+\\
\frac{v+\phi+iz}{\sqrt2}
\end{array}\right],\quad 
\eta =\left[
\begin{array}{c}
\eta_{2/3}\\
\eta_{-1/3}
\end{array}\right],\quad 
\Delta =\left[
\begin{array}{cc}
\frac{\delta_{1/3}}{\sqrt2} & \delta_{4/3} \\
\delta_{-2/3} & -\frac{\delta_{1/3}}{\sqrt2}
\end{array}\right],
\label{component}
\end{align}
where the subscript of the fields represents the electric charge, 
$v \approx 246$ GeV, and $w^\pm$ and $z$ are, respectively, the 
Nambu-Goldstone bosons, which will then be absorbed by the 
longitudinal component of the $W$ and $Z$ bosons.
Due to the $\mu$ term in Eq.~(\ref{Eq:lag-flavor}), the charged components
with $1/3$ and $2/3$ electric charges mix, such that 
their mixing matrices and mass eigenstates are defined as follows: 
\begin{align}
&\left[\begin{array}{c} \eta_{i/3} \\ \delta_{i/3} \end{array}\right] = 
O_i \left[\begin{array}{c} A_i \\ B_i \end{array}\right],\quad
O_i\equiv 
\left[\begin{array}{cc} c_{\alpha_i} & s_{\alpha_i} \\
 -s_{\alpha_i} & c_{\alpha_i}   \end{array}\right], \quad (i=1,2),
\end{align}
where their masses are denoted as $m_{A_i}$ and  $m_{B_i}$ respectively. 
The interactions in terms of the mass eigenstates can be written as 
\begin{align}
& - {L}_{Y}\approx
f_{ij} \overline{ d_{R_i}} \nu_{L_j} (c_{\alpha_1} A_1 +s_{\alpha_1} B_1)
-\frac{g_{ij}}{\sqrt2} \overline{ d_{L_i}^c} \nu_{L_j} (-s_{\alpha_1} A_1 + 
c_{\alpha_1} B_1) 
\label{eq:neut}
\\
&-
f_{ij} \overline{d_{R_i}} \ell_{L_j} (c_{\alpha_2} A_2 +s_{\alpha_2} B_2)
-
\frac{g_{ij}}{\sqrt2} \overline{u_{L_i}^c} \ell_{L_j} (-s_{\alpha_1} A_1 + 
  c_{\alpha_1} B_1) 
\label{eq:lfvs-1}\\
&
-
{g_{ij}} \overline{d_{L_i}^c} \ell_{L_j}\delta_{4/3} \;
 + \; 
  {g_{ij}} \overline{u_{L_i}^c} \nu_{L_j} (-s_{\alpha_2} A_{2}^* + c_{\alpha_2} B_{2}^*) .
\label{eq:lfvs-2}
\end{align}

\begin{figure}[tb]
\begin{center}
\includegraphics[width=80mm]{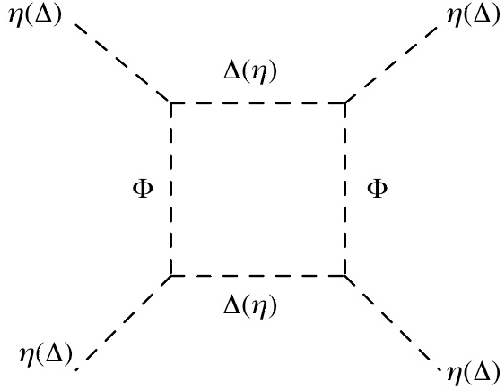}
\caption{ One-loop diagrams for estimating the constraint from vacuum stability.}
\label{fig:VS}
\end{center}
\end{figure}
{\it Vacuum stability}: Since we have charged {components}
such as $\eta_{1/3,2/3}$ and $\delta_{1/3,2/3,4/3}$, 
we have to avoid their pure couplings from becoming negative by 
restricting the negative contribution at one-loop level due to the 
$\mu$ term to be smaller than the tree-level coupling.
{Estimating the one-loop diagrams in Fig.~\ref{fig:VS}}, these conditions are respectively given by
\begin{align}
\frac{\mu^4}{2(4\pi)^2}\int\frac{dxdy\delta(x+y-1)xy}
{(x m^2_\Phi+ym_\delta^2)^2}\lesssim \lambda_\eta^{\rm tree},\quad
\frac{\mu^4}{2(4\pi)^2}\int\frac{dxdy\delta(x+y-1)xy}
{(x m^2_\Phi+ym_\eta^2)^2}\lesssim \lambda_\delta^{\rm tree}, 
\end{align}
where $m_{\eta/\delta}$ are the bare masses in the potential.
{Now we estimate the typical upper bound of $\mu$, assuming $m_{\eta/\delta}\approx1$ TeV that comes from collider {bounds as shall} be seen later. Also, we {restrict} $\lambda_{\eta/\delta}^{\rm tree}\lesssim4\pi$. Under the framework, one obtains $|\mu|\lesssim$ 6.4 TeV, which gives almost no constraint on the TeV scale model.
Thus, we do not need to worry about the stability condition.
}

\begin{figure}[tb]
\begin{center}
\includegraphics[width=80mm]{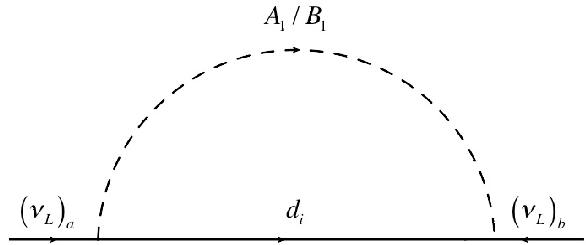}
\caption{ One-loop diagrams for generating the neutrino mass matrix.}
\label{fig:neutrino}
\end{center}
\end{figure}
\subsection{Neutrino mixing}
The dominant contribution to the active neutrino mass matrix $m_\nu$
is given at one-loop level through interactions in
Eq.~(\ref{eq:neut}) {as illustrated in Fig.~\ref{fig:neutrino}}, and its formula is given by
\begin{align}
&(m_{\nu})_{ab}
=
\frac{N_c s_{\alpha_1}c_{\alpha_1} }{2(4\pi)^2}\left[1-\frac{m^2_{A_1}}{m^2_{B_1}}\right]
\sum_{i=1}^3
\left[g^T_{bi} m_{d_i} f_{ia}+ f_{ai} m_{d_i} g_{ib}^T \right] F_I(r_{A_1}, r_{m_{d_i}}),\\
&F_I(r_1, r_2)
=
\frac{r_1(r_2-1)\ln r_1 - r_2(r_1-1)\ln r_2}{(r_1-1)(r_2-1)(r_1-r_2)},\quad (r_1\neq 1),
\end{align}
where $N_c=3$ is the color factor and we define $r_{f}\equiv (m_{f}/m_{B_i})^2$.
$({m}_\nu)_{ab}$ is diagonalized by the Pontecorvo-Maki-Nakagawa-Sakata 
mixing matrix $V_{\rm MNS}$ (PMNS)~\cite{Maki:1962mu} as 
$
({m}_\nu)_{ab} =(V_{\rm MNS} D_\nu V_{\rm MNS}^T)_{ab}$ with $D_\nu\equiv 
(m_{\nu_1},m_{\nu_2},m_{\nu_3})$, where we use the
data in the global analysis~\cite{Forero:2014bxa}.
Then one can parameterize as 
\begin{align}
& g^T R f\equiv \frac12\left[V_{\rm MNS} D_\nu V_{\rm MNS}^T+A\right],\quad
 R\equiv\frac{N_c s_{\alpha_1}c_{\alpha_1} }{2(4\pi)^2}\left[1-\frac{m^2_{A_1}}{m^2_{B_1}}\right]
  \sum_{i=1}^3 m_{d_i} F_I(r_{A_1}, r_{m_{d_i}}),
\end{align}
where $A$ is an arbitrary antisymmetric matrix with complex values.
Finally, we derive the following relations~\cite{Nomura:2016pgg}:
\begin{align}
g
&=\frac12 (V_{\rm MNS}^* D_\nu V_{\rm MNS}^\dag+ A) f^{-1}R^{-1},
\ {\rm or}\
f=\frac12 R^{-1} (g^T)^{-1}(V_{\rm MNS}^* D_\nu V_{\rm MNS}^\dag+ A). 
\end{align}
In the numerical analysis, we shall use the former relation for
convenience.

\subsection{LFVs and FCNCs at tree level}

Leptoquark models often induce LFVs and FCNCs at tree level.
Several processes can arise from the terms $g$ and $f$ and these processes
can be estimated by the effective Hamiltonian~\cite{Carpentier:2010ue} as
\begin{align}
({\cal H}_{\rm eff})_{ijkn}^{\bar\ell\ell\bar d d}
&={f_{kj} f_{in}^\dag }
\left(\frac{c_{\alpha_2}^2} {m_{A_2}^2}+\frac{s_{\alpha_2}^2}{m_{B_2}^2}\right)
(\bar\ell_i\gamma^\mu P_L \ell_j)(\bar d_k\gamma_\mu P_R d_n)
-\frac{g_{kj} g_{in}^\dag }{m_\delta^2}
(\bar\ell_i\gamma^\mu P_L \ell_j)(\bar d_k\gamma_\mu P_L d_n)\nn\\
&\equiv
C_{LR}^{\bar\ell\ell\bar d d} (\bar\ell_i\gamma^\mu P_L \ell_j)(\bar d_k\gamma_\mu P_R d_n)
+
C_{LL}^{\bar\ell\ell\bar d d}(\bar\ell_i\gamma^\mu P_L \ell_j)(\bar d_k\gamma_\mu P_L d_n), \label{eq:hami1}
\\
({\cal H}_{\rm eff})_{ijkn}^{\bar\ell\ell\bar uu}
&=
-\frac{g_{kj} g_{in}^\dag }{2}
\left(\frac{s_{\alpha_1}^2} {m_{A_1}^2}+\frac{c_{\alpha_1}^2}{m_{B_1}^2}\right)
(\bar\ell_i\gamma^\mu P_L \ell_j)(\bar u_k\gamma_\mu P_L u_n)\nn\\
&\equiv C_{LL}^{\bar\ell\ell\bar uu}(\bar\ell_i\gamma^\mu P_L \ell_j)(\bar u_k\gamma_\mu P_L u_n), \label{eq:hami2}
\\
({\cal H}_{\rm eff})_{ijkn}^{\bar\nu\nu\bar qq}
&=
-\frac{g_{kj} g_{in}^\dag }{2}
\left(\frac{s_{\alpha_1}^2} {m_{A_1}^2}+\frac{c_{\alpha_1}^2}{m_{B_1}^2}\right)
(\bar\nu_i\gamma^\mu P_L \nu_j)(\bar q_k\gamma_\mu P_L q_n)\nn\\
&\equiv C_{LL}^{\bar\nu\nu\bar qq}(\bar\nu_i\gamma^\mu P_L \nu_j)(\bar q_k\gamma_\mu P_L q_n).\label{eq:hami3}
\end{align}
Then one can rewrite the relevant coefficients as
\begin{align}
\epsilon_{ijkn}^{\bar\ell\ell\bar d d}&=
\frac{\sqrt2}{4{\rm G_F}} (C_{LR}^{\bar\ell\ell\bar d d}+C_{LL}^{\bar\ell\ell\bar d d}),\\
\epsilon_{ijkn}^{\bar\ell\ell\bar uu}&=
\frac{\sqrt2}{4{\rm G_F}} C_{LL}^{\bar\ell\ell\bar uu},\\
\epsilon_{ijkn}^{\bar\nu\nu\bar qq}&=
\frac{\sqrt2}{4{\rm G_F}} C_{LL}^{\bar\nu\nu\bar qq},
\end{align}
where the experimental bounds on each coefficient are summarized in 
Tables~\ref{tab:lfvs-fcncs-1}, \ref{tab:lfvs-fcncs-2}, 
and \ref{tab:lfvs-fcncs-3}~\cite{Carpentier:2010ue}. 

{\it $B_{d/s}\to\mu^+\mu^-$ measurements}:
Recently, CMS~\cite{Chatrchyan:2013bka} and LHCb~\cite{Aaij:2013aka} 
experiments reported the branching ratios of $B(B_s\to\mu^+\mu^-)$ and 
$B(B_d\to\mu^+\mu^-)$, which can place interesting bounds on new physics.
The bounds on the coefficients of the effective Hamiltonians in
Eq.~(\ref{eq:hami1})~\cite{Sahoo:2015wya} are
\begin{align}
\label{eq:Bsmumu}
& B(B_s\to\mu^+\mu^-):\quad 0\lesssim |C_{LR}^{\bar\mu\mu\bar sb}+C_{LL}^{\bar\mu\mu\bar sb}|\lesssim5\times 10^{-9}\ {\rm GeV}^{-2},\\
\label{eq:Bdmumu}
& B(B_d\to\mu^+\mu^-):\quad 1.5\times 10^{-9} \ {\rm GeV}^{-2}
\lesssim |C_{LR}^{\bar\mu\mu\bar db}+C_{LL}^{\bar\mu\mu\bar db}|\lesssim3.9\times 10^{-9}\ {\rm GeV}^{-2},
\end{align}
where the phase is assumed to be zero for simplicity. 
The bounds from the other modes are
\begin{align}
& B(B_s\to e^+ e^-):\quad 
 |C_{LR}^{\bar ee \bar sb}+C_{LL}^{\bar ee\bar sb}|\lesssim2.54\times 10^{-5}\ {\rm GeV}^{-2},\\
& B(B_d\to e^+ e^-):\quad 
 |C_{LR}^{\bar ee\bar db}+C_{LL}^{\bar ee\bar db }|\lesssim1.73\times 10^{-5}\ {\rm GeV}^{-2},\\
 & B(B_s\to \tau^+\tau^-):\quad 
 |C_{LR}^{\bar \tau\tau \bar sb}+C_{LL}^{\bar \tau\tau \bar sb}|\lesssim1.2\times 10^{-8}\ {\rm GeV}^{-2},\\
& B(B_d\to \tau^+ \tau^-):\quad 
 |C_{LR}^{\bar \tau\tau\bar db}+C_{LL}^{\bar \tau\tau\bar db}|\lesssim1.28\times 10^{-6}\ {\rm GeV}^{-2}.
\end{align}

 \begin{table}[t]
\begin{tabular}{|c|c|c|c|c} \hline
${ijkn}$ of $\epsilon_{ijkn}^{\bar\ell\ell\bar d d}$ & Constraints on $\epsilon_{ijkn}^{\bar\ell\ell\bar d d}$ & Observable & Experimental value \\ \hline\hline
$eeds(\to1112)$ & $5.7\times 10^{-5}$ & $B(K^0_L\to \bar ee)$ & $9.0\times 10^{-12}$   \\
$eedb(\to1113)$ & $2.0\times 10^{-4}$ & $\frac{B(B^+\to \pi^+\bar ee)}{B(B^0\to \pi^-\bar e\nu_e)}$ &$<\frac{1.8\times 10^{-7}}{1.34\times 10^{-4}}$   \\
$eesb(\to1123)$ & $1.8\times 10^{-4}$ & $\frac{B(B^+\to K^+\bar ee)}{B(B^0\to D^0\bar e\nu_e)}$ & $\frac{4.9\times 10^{-7}}{2.2\times 10^{-2}}$  \\ 
$e\mu dd(\to1211)$ & $8.5\times 10^{-7}$ & $\mu-e$ conversion on Ti & $\frac{\sigma(\mu^-Ti\to e^-Ti)}{\sigma(\mu^-Ti\to capture)}<4.3\times 10^{12}$  \\
$e\mu ds(\to1212)$ & $3.0\times 10^{-7}$ & $B(K^0_L\to \bar e\mu)$ & $<4.7\times10^{-12}$  \\ 
$e\mu db(\to1213)$ & $2.0\times 10^{-4}$ & $\frac{B(B^+\to \pi^+\bar e\mu)}{B(B^0\to \pi^-\bar e\nu_e)}$ & $\frac{1.7\times 10^{-7}}{1.34\times 10^{-4}}$  \\ 
$e\mu sb(\to1223)$ & $8\times 10^{-5}$ & $\frac{B(B^+\to K^+\bar e\mu)}{B(B^+\to D^0\bar e\nu_e)}$ & $<\frac{9.1\times 10^{-8}}{2.2\times 10^{-2}}$  \\ 
$e\tau dd(\to1311)$ & $8.4\times 10^{-4}$ & $\frac{B(\tau\to \pi^0 e)}{B(\tau\to \pi\nu_\tau)}$ & $<\frac{8\times 10^{-8}}{10.91\times 10^{-2}}$  \\ 
$e\tau ds(\to1312)$ & $4.9\times 10^{-4}$ & $\frac{B(\tau\to eK)}{B(\tau\to \bar\nu K)}$ & $B<3.3\times10^{-8}$  \\ 
$e\tau db(\to1313)$ & $4.1\times 10^{-3}$ & $B(B^0\to \bar e\tau)$ & $<1.1\times10^{-4}$  \\ 
$\mu\mu ds(\to2212)$ & $7.8\times 10^{-6}$ & $B(K_L^0\to \bar\mu\mu)$ & $6.84\times10^{-9}$  \\ 
$\mu\mu db(\to2213)$ & $1.3\times 10^{-4}$ & $\frac{B(B^+\to\pi^+\bar\mu\mu)}{B(B^0\to\pi^- \bar e\nu_e)}$ & $<\frac{6.9\times10^{-8}}{1.3\times 10^{-4}}$ \\ 
$\mu\tau dd(\to2311)$ & $9.8\times 10^{-4}$ & $\frac{B(\tau\to\pi^0\mu)}{B(\tau\to\pi^- \nu_\tau)}$ & $<\frac{1.1\times10^{-7}}{10.91\times 10^{-2}}$  \\ 
$\mu\tau ds(\to2312)$ & $5.4\times 10^{-4}$ & $\frac{B(\tau\to\mu K)}{B(\tau\to \bar\nu K)}$ & $B<4.0\times10^{-8}$  \\ 
$\mu\tau db(\to2313)$ & $2.1\times 10^{-2}$ & $B(B^0\to \bar \mu\tau)$ & $<2.2\times10^{-5}$  \\ 
$\mu\tau sb(\to2323)$ & $2.3\times 10^{-3}$ & $\frac{B(B^+\to K^+\bar\tau \mu)}{B(B^+\to D^0\bar e \nu)}$ & $<\frac{7.7\times10^{-5}}{2.2\times10^{-2}}$  \\ 
$\tau\tau db(\to3313)$ & $0.2$ & $B(B^0\to \bar\tau\tau)$ & $<4.1\times10^{-3}$\\ 
 \hline
\end{tabular}
\caption{
Summary for the experimental bounds on $\epsilon_{ijkn}^{\bar\ell\ell\bar d d}$.}
\label{tab:lfvs-fcncs-1}
\end{table}

 \begin{table}[t]
\begin{tabular}{|c|c|c|c|c} \hline
${ijkn}$ of $\epsilon_{ijkn}^{\bar\ell\ell\bar uu}$ & Constraints on $\epsilon_{ijkn}^{\bar\ell\ell\bar uu}$ & Observable & Experimental value \\ \hline\hline
$eeuc(\to1112)$ & $7.9\times 10^{-3}$ & $\frac{B(D^+\to\pi^+\bar ee)}{B(D^0\to\pi^-\bar e\nu_e)}$ & $<\frac{7.4\times 10^{-6}}{2.83\times10^{-3}}$   \\
$eett(\to1133)$ & $0.092$ & $Z\to\bar ee$ & $R_e=20.804\pm 0.050$   \\
$e\mu uu(\to1211)$ & $8.5\times10^{-7}$ & $\mu-e$ conversion on Ti & $\frac{\sigma(\mu Ti\to eTi)}{\sigma(\mu Ti\to capture)}<4.3\times 10^{-12}$   \\
$e\mu uc(\to1212)$ & $1.7\times10^{-2}$ & $\frac{B(D^+\to \pi^+\bar e\mu)}{B(D^0\to\pi^-\bar e\nu_e)}$ & $<\frac{3.4\times 10^{-5}}{2.83\times10^{-3}}$   \\
$e\mu tt(\to1233)$ & $0.1$ & $Z\to \bar e\mu$ & $B<1.7\times10^{-6}$   \\
$e\tau uu(\to1311)$ & $8.4\times10^{-4}$ & $\frac{B(\tau\to \pi^0 e)}{B(\tau\to\pi^-\nu_\tau)}$ & $<\frac{8\times 10^{-8}}{10.91\times10^{-2}}$   \\
$e\tau tt(\to1233)$ & $0.2$ & $Z\to \bar e\tau$ & $B<9.8\times10^{-6}$   \\
$\mu\mu uc(\to2212)$ & $6.1\times10^{-3}$ & $\frac{B(D^+\to \pi^+ \bar\mu\mu)}{B(D^0\to\pi^-\bar e\nu_e)}$ & $<\frac{3.9\times 10^{-6}}{2.83\times10^{-3}}$   \\
$\mu\mu tt(\to2233)$ & $0.061$ & $Z\to \bar \mu\mu$ & $R_\mu=20.785\pm0.033$   \\
$\mu\tau uu(\to2311)$ & $9.8\times10^{-4}$ & $\frac{B(\tau\to \pi^0 \mu)}{B(\tau\to\pi^-\nu_\tau)}$ & $<\frac{1.1\times 10^{-7}}{10.91\times10^{-2}}$   \\
$\mu\tau tt(\to2333)$ & $1$ & $Z\to \tau\bar\mu$ & $B<12\times10^{-6}$   \\
$\tau\tau tt(\to3333)$ & $0.086$ & $Z\to \bar \tau\tau$ & $R_\tau=20.764\pm0.045$   \\
 \hline
\end{tabular}
\caption{
Summary for the experimental bounds on $\epsilon_{ijkn}^{\bar\ell\ell\bar uu}$.}
\label{tab:lfvs-fcncs-2}
\end{table}

 \begin{table}[t]
\begin{tabular}{|c|c|c|c|c} \hline
${ijkn}$ of $\epsilon_{ijkn}^{\bar\nu_i\nu_j\bar q_kq_n}$ & Constraints on $\epsilon_{ijkn}^{\bar\nu_i\nu_j\bar q_kq_n}$ & Observable & Experimental value \\ \hline\hline
$ijds(\to ij12)$ & $9.4\times 10^{-6}$ & $\frac{B(K^+\to\pi^+\bar \nu\nu)}{B(K^+\to\pi^0\bar e\nu_e)}$ & $\frac{1.5\times 10^{-10}}{5.08\times10^{-2}}$   \\ \hline
\end{tabular}
\caption{Summary for the experimental bounds on $\epsilon_{ijkn}^{\bar\nu_i\nu_j\bar q_kq_n}$, where $(i,j)=(1-3)$.}
\label{tab:lfvs-fcncs-3}
\end{table}

\begin{figure}[tb]
\begin{center}
\includegraphics[width=80mm]{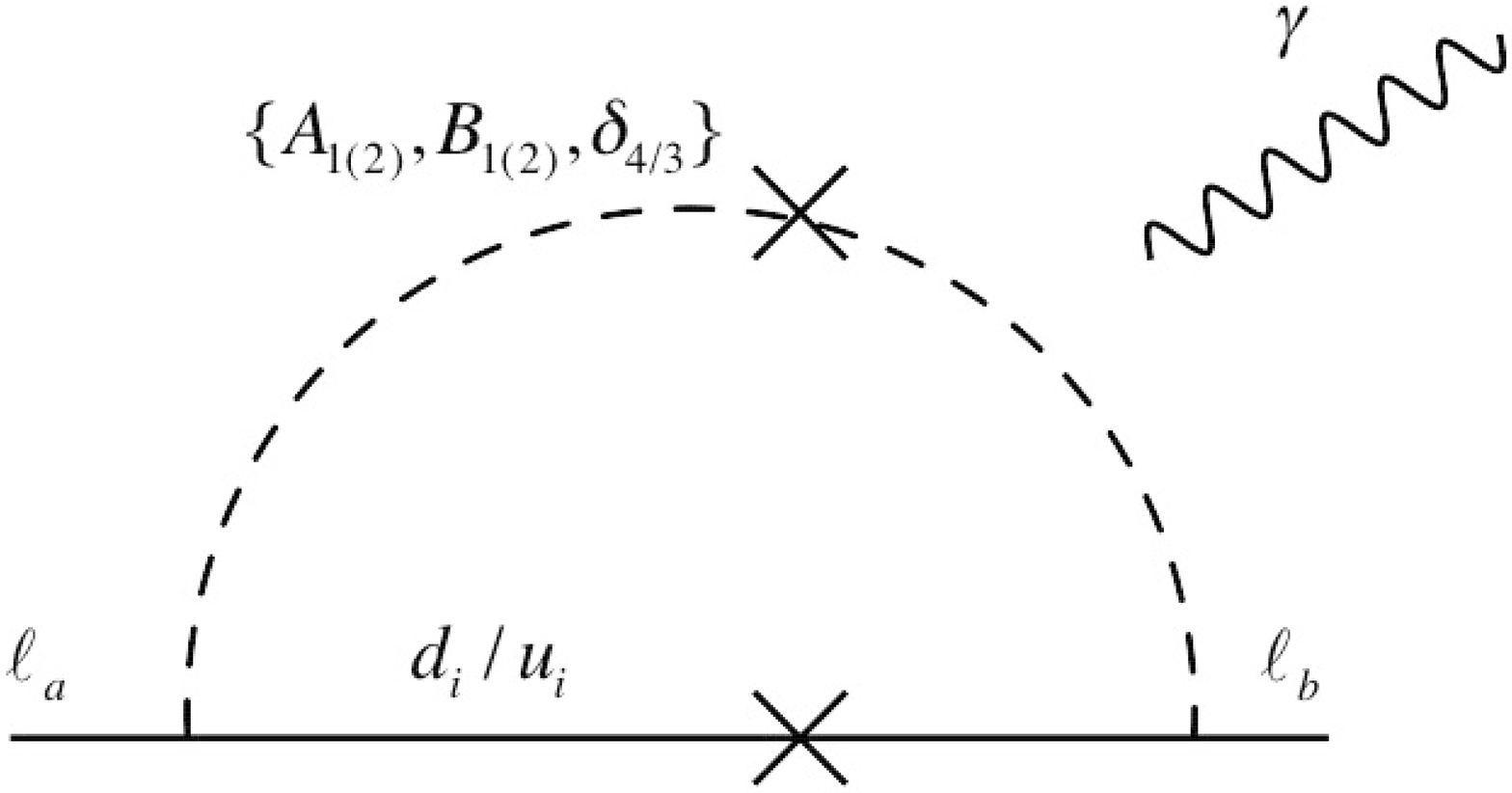}
\caption{ One-loop diagrams for the LFV processes $\ell_a \to \ell_b \gamma$ 
where the cross mark on the internal lines indicates the  attachment of a
photon line.}
\label{fig:LFV}
\end{center}
\end{figure}
\subsection{LFVs and FCNCs at the one-loop level}
\label{lfv-lu}
{\it LFVs}:
$\ell_a\to\ell_b\gamma$ processes, which arise from Eqs.~(\ref{eq:lfvs-1}) 
and ~(\ref{eq:lfvs-2}) via one-loop diagrams {as shown in Fig.~\ref{fig:LFV}}, often give 
stringent experimental constraints, and the branching ratio is given by
\begin{align}
B(\ell_a\to\ell_b \gamma)
=
\frac{48\pi^3 C_a\alpha_{\rm em}}{{\rm G_F^2} m_a^2 }(|(a_R)_{ab}|^2+|(a_L)_{ab}|^2),
\end{align}
where $m_{a(b)}$ is the mass for the charged-lepton eigenstate, 
$C_{a}=(1,1/5)$ for ($a=\mu,\tau$).
$a_L$ and $a_R$ are respectively given by
\begin{align}
&\frac{(a_R)_{ab}}{N_c}\approx
-\frac{ f^\dag_{bi} f_{ia} m_a }{12(4\pi)^2}
\left( \left[\frac{Q_{A_2} c^2_{\alpha_2} }{m_{A_2}^2} + \frac{Q_{B_2} s^2_{\alpha_2} }{m_{B_2}^2}\right]
+
2 \left[\frac{Q_d  c^2_{\alpha_2} }{m_{A_2}^2} + \frac{Q_d  s^2_{\alpha_2} }{m_{B_2}^2}\right]\right)
\\
&
+\frac{ g^\dag_{bi} g_{ia} {m_a} }{24(4\pi)^2} 
\left(
 \left[\frac{- Q_{\delta} }{m_\delta^2}+ \frac{2Q_{\bar d}}{m_\delta^2}\right]
-
 \left[\frac{Q_{A_1} s^2_{\alpha_1} }{m_{A_1}}+ \frac{Q_{B_1}  c^2_{\alpha_1} }{m_{B_1}}\right]
+2 Q_{\bar u}
 \left[\frac{s^2_{\alpha_1} }{m_{A_1}}+ \frac{c^2_{\alpha_1}}{m_{B_1}}\right]\right),\nn
\end{align} 
where we have assumed $m_{d(u)}<<m_{A_i,B_i,\delta}(i=1-2)$, and 
$a_L=a_R(m_a\rightarrow m_b)$, 
$Q_{\delta}=-Q_{\bar\delta}\equiv -4/3$, $Q_{A_2}=Q_{B_2}\equiv -2/3$, 
$Q_{A_1}=Q_{B_1}\equiv -1/3$, $Q_{d}=-Q_{\bar d}\equiv -1/3$, 
$Q_{u}=-Q_{\bar u}\equiv 2/3$.
The current experimental upper bounds are given 
by~\cite{TheMEG:2016wtm, Adam:2013mnn}
  \begin{align}
  B(\mu\rightarrow e\gamma) &\leq4.2\times10^{-13},\quad 
  B(\tau\rightarrow \mu\gamma)\leq4.4\times10^{-8}, \quad  
  B(\tau\rightarrow e\gamma) \leq3.3\times10^{-8}~.
 \label{expLFV}
 \end{align}

{\it The muon anomalous magnetic moment (muon $g-2$)}: It 
has been measured with 
a high precision, and its deviation from the SM prediction is of 
order $10^{-9}$. The formula for the muon $g-2$ is given by
\begin{align}
\Delta a_\mu\approx -m_\mu [{(a_R)_{22}+(a_L)_{22}}].\label{damu}
\end{align}
In our model, typical values are at most $10^{-14}$ with mostly the negative 
sign.  Although there may exist some positive sources, however, 
the negative sources are always larger than the positive ones 
when we impose all the bounds from LFVs and FCNCs.  

{\it $Q-\overline{Q}$ mixing}: We also consider the constraint of the 
$Q-\overline{Q}$ mixing, where $Q=K,B,D$.
The mixing is characterized by $\Delta m_Q$ given by~\cite{hep-ph/9604387}
{
\begin{align}
\Delta m_K&\approx
\sum_{i,j=1}^3\frac{
5f_K^2 m_K^3\left(
8f_{i2}^\dag f_{1i} f_{1j} f^\dag_{j2} m_{B_1}^2 m_{\delta}^2
+
g_{2i} g^\dag_{i1} g_{2j} g^\dag_{j1} (4 m_{B_1}^2 + m_{\delta}^2)\right)}
{768 \pi^2 m_{A_1}^2 m_{B_1}^2 m_{\delta}^2 (m_d+m_s)^2}
 \lesssim 3.48\times10^{-15}[{\rm GeV}],\label{eq:kk}\\
\Delta m_B&\approx
\sum_{i,j=1}^3\frac{
5f_B^2 m_B^3\left(
8f_{i2}^\dag f_{3i} f_{3j} f^\dag_{j2} m_{B_1}^2 m_{\delta}^2
+
g_{2i} g^\dag_{i3} g_{2j} g^\dag_{j3} (4 m_{B_1}^2 + m_{\delta}^2)\right)}
{768 \pi^2 m_{A_1}^2 m_{B_1}^2 m_{\delta}^2 (m_b+m_s)^2}
  \lesssim 3.36\times10^{-13} [{\rm GeV}],\label{eq:bb}\\
\Delta m_D&\approx
\sum_{i,j=1}^3\frac{
5f_D^2 m_D^3g_{2i} g^\dag_{i1} g_{2j} g^\dag_{j1} }
{768 \pi^2 m_{B_1}^2  (m_u+m_c)^2}
 \lesssim 6.25\times10^{-15}[{\rm GeV}],\label{eq:bb},
\end{align}
where we assume $m_{A_1}\approx m_{A_2}$, $m_{B_1}\approx m_{B_2}$, and $s_{\alpha_{1(2)}}\approx0$, and  
each of the last inequalities of Eqs.(\ref{eq:kk}, \ref{eq:bb})
represents the upper bound on the experimental values \cite{pdg}, and
$f_K\approx0.156$ GeV, $f_B\approx0.191$ GeV, $m_K\approx0.498$ GeV,
and $m_B\approx5.280$ GeV.
}

{\it $b\to s\gamma$}: It can arise from the same term in the LFVs, yet 
the constraint is always weaker than those of LFVs. Thus we do not further 
consider this process.

\subsection{ Oblique parameters} 
Since $\eta$ and $\Delta$ are multiplets under the $SU(2)_L$ gauge symmetry, 
we need to take into account the constraints from the oblique parameters 
$S$, $T$, and $U$. 
Here we focus on the new physics contributions to $S$ and $T$ parameters, 
$\Delta S$ and $\Delta T$, which are defined by
\begin{align}
\Delta S&={16\pi} \frac{d}{dq^2}[\Pi_{33}(q^2)-\Pi_{3Q}(q^2)]|_{q^2\to0},\quad
\Delta T=\frac{16\pi}{s_{W}^2 m_Z^2}[\Pi_{\pm}(0)-\Pi_{33}(0)],
\end{align}
where $s_{W}^2\approx0.22$ is the Weinberg angle and $m_Z$ is the $Z$ 
boson mass. 
The loop factors $\Pi_{33,3Q,QQ,\pm}(q^2)$ are calculated from the one-loop 
vacuum-polarization diagrams for $Z$ and $W^\pm$ bosons, 
$i \Pi_{Z(W)}^{\mu \nu}$, where new particles run inside the loop diagrams, 
as follows;
\begin{align}
\label{eq:piZ}
\Pi_{Z}^{\mu \nu} &=  g^{\mu \nu} \frac{e^2}{c_W^2 s_W^2} \left( \Pi_{33}(q^2) - 2 s_W^2 \Pi_{3Q}(q^2) - s_W^4 \Pi_{QQ}(q^2) \right), \\
\label{eq:piW}
\Pi_{W}^{\mu \nu} &= g^{\mu \nu} \frac{e^2}{s_W^2} \Pi_{\pm}(q^2), 
\end{align}
The list of new particle contributions is quite lengthy and so we 
summarize them in the Appendix.
The experimental bounds are given by \cite{pdg}
\begin{align}
(0.05 - 0.09) \le \Delta S \le (0.05 + 0.09), \quad 
(0.08 - 0.07) \le \Delta T \le (0.08 + 0.07).
 \end{align}

\subsection{Collider physics}

The interactions of the $\eta$ and $\Delta$ are very similar to leptoquarks
or squarks.  The first signature that we consider is their effects on 
Drell-Yan production and also the lepton-flavor violating production
processes such as $e^\pm \mu^\mp$, $\mu^\pm \tau^\mp$, and $e^\pm \tau^\mp$ {\cite{Mandal:2015lca}}.

Without loss of generality we take the mixing angles between $\eta$ and 
$\Delta$ to be small (indeed required by the $S,T$ parameters), such
that $\eta_{1/3,2/3} \approx A_{1,2}$ and $\delta_{1/3,2/3} \approx B_{1,2}$.
We can write down the amplitude for 
$d_{R_i} (p_1)  \overline{d_{R_{i'}}} (p_2) \to \ell_{L_j} (q_1) \bar 
\ell_{L_{j'}} (q_2) $ with a $t$-channel exchange of $\eta_{2/3}$
\begin{eqnarray}
i {\cal M} &=& - i f_{ij} f_{i'j'} \frac{1}{\hat t - m_{\eta}^2} 
 \bar u (q_1) P_R u(p_1 ) \; \bar v(p_2) P_L v(q_2) \nonumber \\
 &  \stackrel{Fierz}{ = }  &
  - i f_{ij} f_{i'j'} \frac{1}{\hat t - m_{\eta}^2}  \frac{1}{2}
  \bar u (q_1) \gamma^\mu P_L  v (q_2 ) \; \bar v(p_2) \gamma_\mu P_R u(p_1)  \;.
\end{eqnarray}
When $|\hat t| \ll m^2_{\eta}$ we can identify this amplitude as a 4-fermion
contact interaction and equate
\begin{equation}
   \frac{f_{ij} f_{i'j'} }{2 m_{\eta}^2}  = \frac{4\pi}{\Lambda^2_{LR}}
\end{equation}
where $\Lambda_{LR}$, with $L\,(R) $ chirality refers to the lepton (quark),
is often the limit quoted for the 4-fermion contact interactions.
Since only the limits $\Lambda_{LL}$ are quoted in PDG \cite{pdg}, which
was based on Ref.~\cite{cheung}, we use the limit of $\Lambda_{LR}$ obtained
in Ref.~\cite{cheung}. The limit on $\Lambda_{LR} \approx 11 -16$ TeV depending
on the sign of the 4-fermion contact interaction. Let us simply take 
$\Lambda_{LR} = 16$ TeV, and translate into the mass limit of $m_\eta$ as
follows (with $i=i'=1$ and $j=j'=1$ or 2)~\footnote{
{The most updated limits by the ATLAS and CMS on the compositeness 
scale are $\Lambda_{\pm}(LL) \agt 17-25$ TeV (ATLAS) \cite{atlas-contact}
and $11-18$ TeV (CMS) \cite{cms-contact}, 
which are somewhat less restrictive 
than the limits that we quoted from the PDG.  We therefore used the PDG values.
Nevertheless, the limits are not as stringent as the direct search limits of 
around 1 TeV provided that the values for $f_{1j}$ and $g_{1j}$ are less $O(10^{-1})$.  
}}
\begin{equation}
\label{eta-limit}
 m_{\eta} \agt f_{1j} \times 3.2 \; {\rm TeV}  \qquad (j=1,2) \;.
\end{equation}
{The effect of including the $\hat t$ or $\hat u$ in the leptoquark propagator
has been explicitly worked out in Refs.~\cite{1409.2372,1410.4798}. It was shown
that the limits obtained with the proper leptoquark propagators are weakened 
by about 40\% to a few \% for leptoqark mass of 1 TeV to 3 TeV. Nevertheless,
the direct search limits of around 1 TeV are more restrictive then.
}

Note that the approximation $1/(\hat t - m_\eta^2) \simeq 1/(- m_\eta^2)$ may 
not be valid for $m_\eta \alt 1$ TeV. Yet, the limit obtained in 
Eq.~(\ref{eta-limit}) is a rough estimate on how heavy the $\eta$ boson can
be without upsetting the current Drell-Yan data.  If the $\eta$ boson
is around 1 TeV, the Drell-Yan invariant-mass distribution
may receive some enhancement at the large invariant-mass end.  

Similarly, we can write down the amplitudes for 
$ u^c_{L_i} \overline{ u^c_{L_{i'}} } \to \ell_{L_j} \bar \ell_{L_{j'}}$ and
$ d^c_{L_i} \overline{ d^c_{L_{i'}} }\to \ell_{L_j} \bar \ell_{L_{j'}}$ 
with the exchange
of $\delta_{1/3}$ and $\delta_{4/3}$, respectively. 
The resulting mass limits on $m_{\delta}$ can be written as
\begin{equation}
  \frac{g_{1j} g_{1j} }{2 m_\delta^2} = \frac{4 \pi}{\Lambda_{LL}^2} \;.
\end{equation}
With a more severe $\Lambda_{LL} \approx 25$ TeV, we obtain
\begin{equation}
\label{delta-limit}
 m_{\delta} \agt g_{1j} \times 5.0 \;{\rm TeV} \qquad (j=1,2) \;.
\end{equation}
We observe that the mass limit on $\delta$ is somewhat stronger
than $\eta$, simply because of the chiralities of quarks and leptons
that they induce.

On the other hand, the $\eta$ and $\delta$ bosons can be directly pair
produced by the strong interaction, followed by their decays into
leptons and quarks. Therefore, the typical signature would be a pair
of leptons and a pair of jets in the final states, of which the invariant
mass of one jet and one lepton shows a clear peak.
Note that the jets can be light or heavy flavors depending on the 
Yukawa couplings $f_{ij}$ and $g_{ij}$, and the leptons can be neutrinos or 
charged leptons of different or same flavors. 
The current limits on leptoquarks, using electron or muon plus jets, are
about 1 TeV \cite{lq-limit}. 
Pair production cross sections have been calculated with NLO accuracy
in Ref.~\cite{pair} long time ago.  The cross section at 13 TeV LHC is
of order $O(10)$ fb for 1 TeV $\eta$ or $\delta$ boson.
Combining the direct search limit of about 1 TeV for $\eta$ and $\delta$,
and Eqs.~(\ref{eta-limit}) and (\ref{delta-limit}), we obtain upper limits
for $f_{1j}$ and $g_{1j}$:
\begin{equation}
  f_{1j} \alt 0.3, \qquad g_{1j} \alt 0.2 \qquad (j=1,2) \;.
\end{equation}
{A list of more comprehensive collider and low energy constraints can be 
found in Ref.~\cite{1603.04993}. }

\section{ $b\to s\bar \ell \ell$ Anomaly and Predictions}  
The more striking anomaly was the lepton-universality violation measured 
in the ratio
$R_K \equiv B(B\to K\mu\mu)/B(B\to K ee) = 0.745 ^{+0.090}_{-0.074} \pm 0.036$
by LHCb \cite{lhcb-2014}, and the less one was the angular distributions
of $B \to K^*\mu\mu$~\cite{lhcb-2013}.
The new-physics contributions to the effective Hamiltonian 
characterizing the decay processes are
\begin{align}
{\cal H}_{\rm eff}^f&=\frac{f_{b\ell} f_{s\ell'} }4
\left(\frac{c_{\alpha_2}^2} {m_{A_2}^2}+\frac{s_{\alpha_2}^2}{m_{B_2}^2}\right)
\left[(\bar s\gamma^\mu P_R b)(\bar \ell'\gamma_\mu \ell)-(\bar s\gamma^\mu P_R b)(\bar \ell'\gamma_\mu\gamma_5 \ell) \right],
\\
{\cal H}_{\rm eff}^g&=-\frac{g_{b\ell} g_{s\ell'} }{4m_{\delta}^2}
\left[(\bar s\gamma^\mu P_L b)(\bar \ell'\gamma_\mu \ell)  
{-}
(\bar s\gamma^\mu P_L b)(\bar \ell'\gamma_\mu\gamma_5 \ell) \right].
\end{align}
Therefore, the relevant new-physics Wilson coefficients for the operators
$C^{(')}_{9,10}$ are given by
\begin{align}
(C_9')^{\ell \ell'}&=\frac{1}{C_{\rm SM}}\frac{f_{b\ell} f_{s\ell'}}{4}\left(\frac{c_{\alpha_2}^2} {m_{A_2}^2}+\frac{s_{\alpha_2}^2}{m_{B_2}^2}\right),\quad
(C_{10}' )^{\ell \ell'}=- (C_9' )^{\ell \ell'},\\
(C_9)^{\ell \ell'}&={-}(C_{10})^{\ell \ell'}=-\frac{1}{C_{\rm SM}}\frac{g_{b\ell} g_{s\ell'} }{4m_{\delta}^2},\quad C_{\rm SM}\equiv \frac{V_{tb}V_{ts}^* G_F\alpha_{\rm em}}{\sqrt2\pi},
\end{align}
where $\alpha_{\rm em}\approx1/137$ is the fine-structure constant,
and ${\rm G_F}\approx1.17\times 10^{-5}$ GeV$^{-2}$ is the Fermi
constant. In our analysis, we focus on the case of $\ell=\ell' = \mu$
and we write
the $\mu \mu$ component simply as $C_9(C_{10})$ and $C_9'(C_{10}')$ in the
following. In Table~\ref{tab:C910}~\cite{Descotes-Genon:2015uva}, we
summarize the best fit values of the Wilson coefficients for
explaining the experimental anomalies where we focus on the cases of
$C_9 = {-}C_{10}$ and $C_9' = - C_{10}'$ since most of the allowed
parameter sets provide either $C_9 \ll C_9'$ or $C_9 \gg C_9'$ as we
show in our numerical analysis.
{Note that $(C_9)^{\mu \mu}$ and $(C'_9)^{ \mu\mu}$ are roughly estimated as 
\begin{equation}
C_9[C_9'] \sim 3.3 \times 10^2 \left( \frac{1 \, {\rm TeV}}{m_{LQ}} \right)^2 g_{b \mu} g_{s \mu} [f_{b \mu} f_{s \mu}],
\end{equation}
where $m_{LQ}$ indicates the leptoquark mass. Therefore, the bi-product of the couplings are required to be $\sim O(10^{-3})$ to obtain the best fit value for $\sim$1 TeV leptoquark mass.
}
\begin{table}[t]
\begin{tabular}{c|c|c|c} \hline
 & ${\rm Best\ fit}$ & $1\sigma$ & $3\sigma$ \\ \hline
$C_{9}={-}C_{10}$ & $-{0.68}$ & $[{-0.85,-0.50}]$& $[{-1.22,-0.18}]$ \\
$C'_{9}=-C'_{10}$ & $0.19$ & $[0.07,0.31]$& $[-0.17,0.55]$ \\ \hline
\end{tabular}
\caption{Summary for the new physics contribution to $C_{9,10}(C'_{9,10})$ explaining experimental anomalies of the $b\to s\bar \ell \ell$ processes
in the cases of $C_9 = - C_{10}$ and $C'_9 = - C'_{10}$ where new physics contribution is nonzero only for the values on the table for both cases. }
\label{tab:C910}
\end{table}

\subsection{Numerical analysis \label{sec:numerical}}

We are now ready to search for the allowed parameter space, which
satisfies all the constraints that we have discussed above, 
in particular those explaining the $b\to s \mu \mu$ anomalies.
First of all, we fix some mass parameters as $m_{A_2}=m_{A_1}$ 
and $m_{B_2}=m_{\delta}=m_{B_1}$ {where we require degenerate masses for the components of $\eta$ and $\Delta$ to suppress the oblique parameters $\Delta S$ and $\Delta T$}.
We prepare 80 million random sampling points for the relevant input 
parameters with the following ranges:
{
\begin{align}
& (m_{A_1},m_{B_1}) \in [1\,, 5\,]\text{TeV},\quad
|A_{12,23,13}| \in \left[10^{-13},\ 10^{-7}\,\right] \text{GeV}, \nn\\
& (\alpha_{1} ,\  \alpha_{2} ) \in [10^{-5}, 10^{-2}], \quad
| f_{ij} | \in [10^{-5}, {4 \pi}].
\label{range_scanning}
\end{align}
}
After scanning, we find {709} parameter sets, which can fit 
neutrino oscillation data and satisfy all the constraints.
Note that mixing angle $\alpha_{1(2)}$ is required to be small 
due to the constraints from $\Delta S$ and $\Delta T$ parameters.

\begin{figure}[tb]
\begin{center}
\includegraphics[width=80mm]{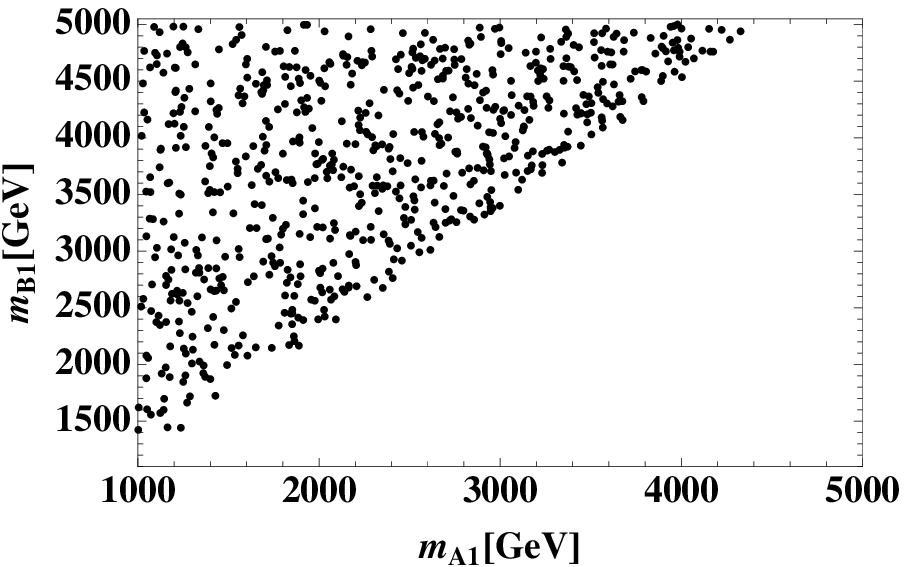}
\includegraphics[width=80mm]{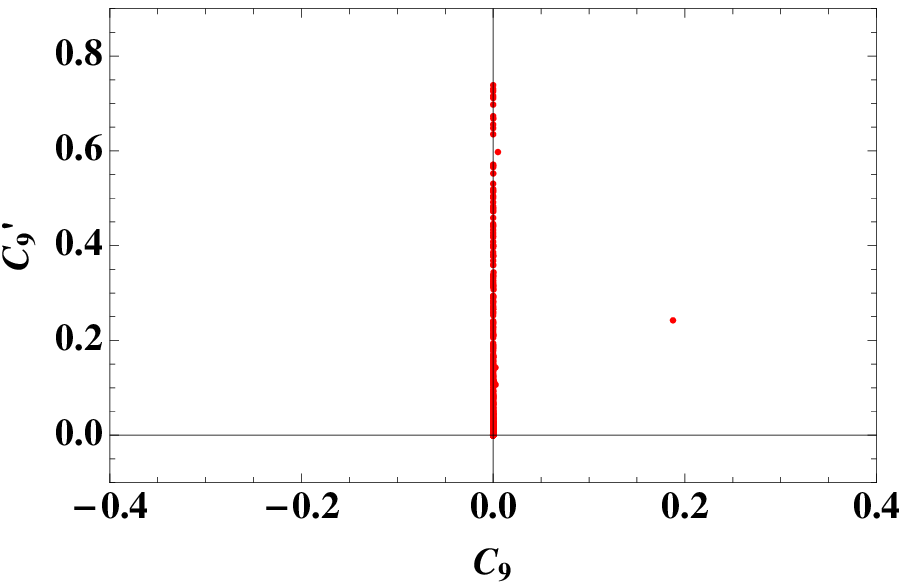}
\caption{
Scattering {\it} plots in the plane of $m_{A_1}$ versus $m_{B_1}$ in the left
panel; and in the plane of $C_9$ versus $C_{9}'$ in the right panel. 
{
The vertical line of 1 TeV is superimposed in the left panel due to the 
direct search limit of leptoquarks.}}
\label{fig:nums}
\end{center}
\end{figure}

\begin{figure}[t!]
\includegraphics[width=80mm]{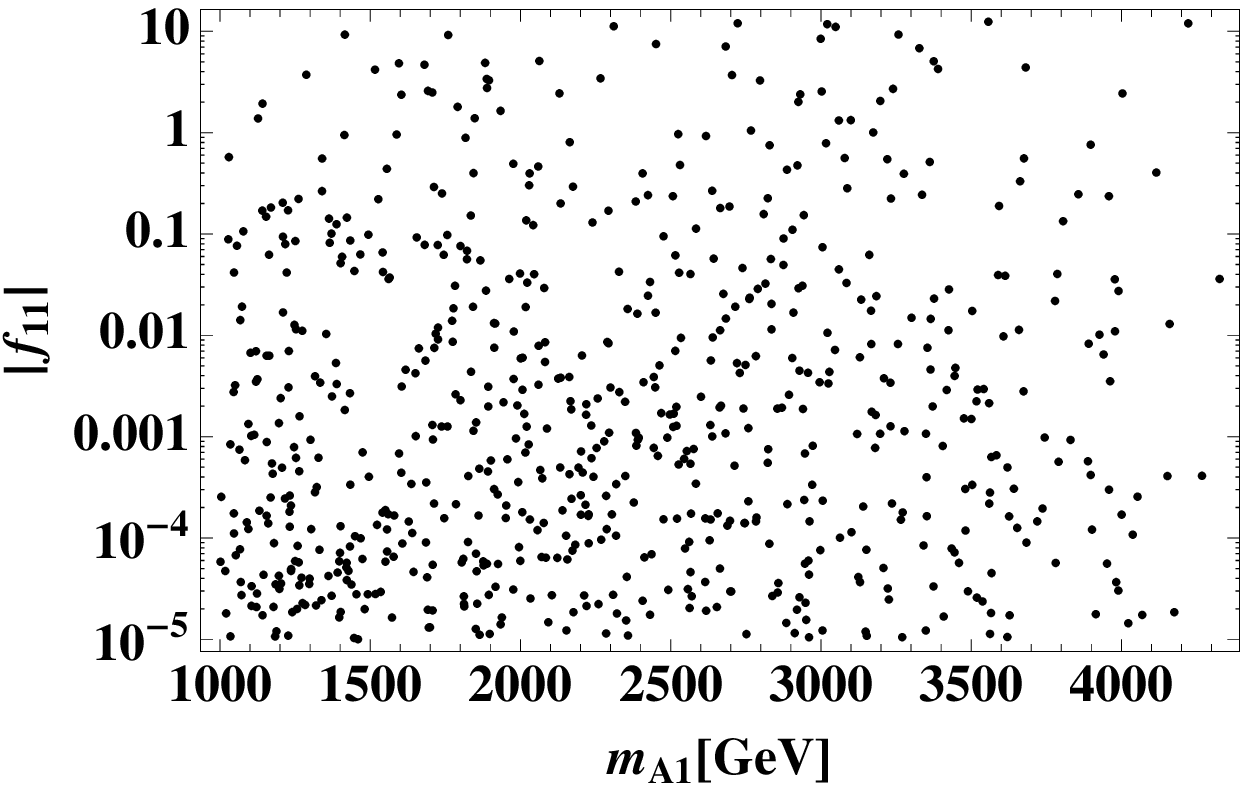}
\includegraphics[width=80mm]{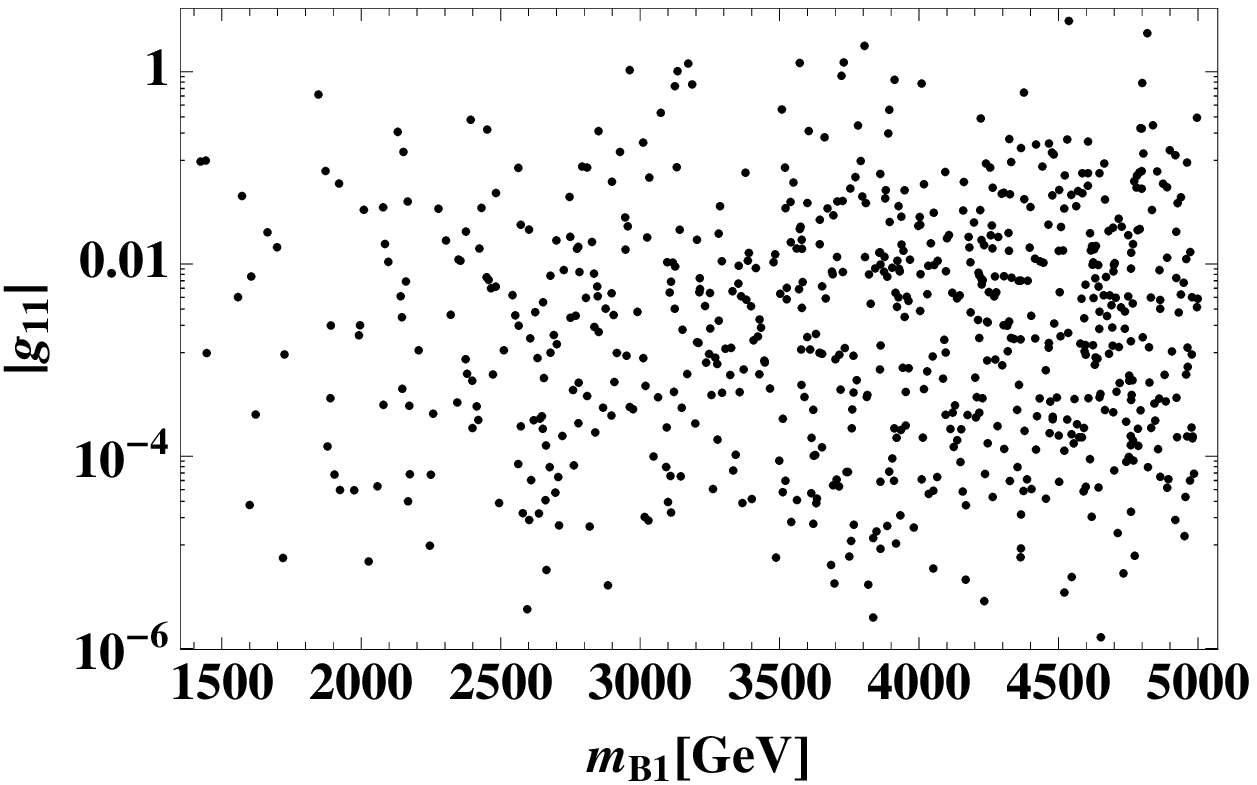}
\includegraphics[width=80mm]{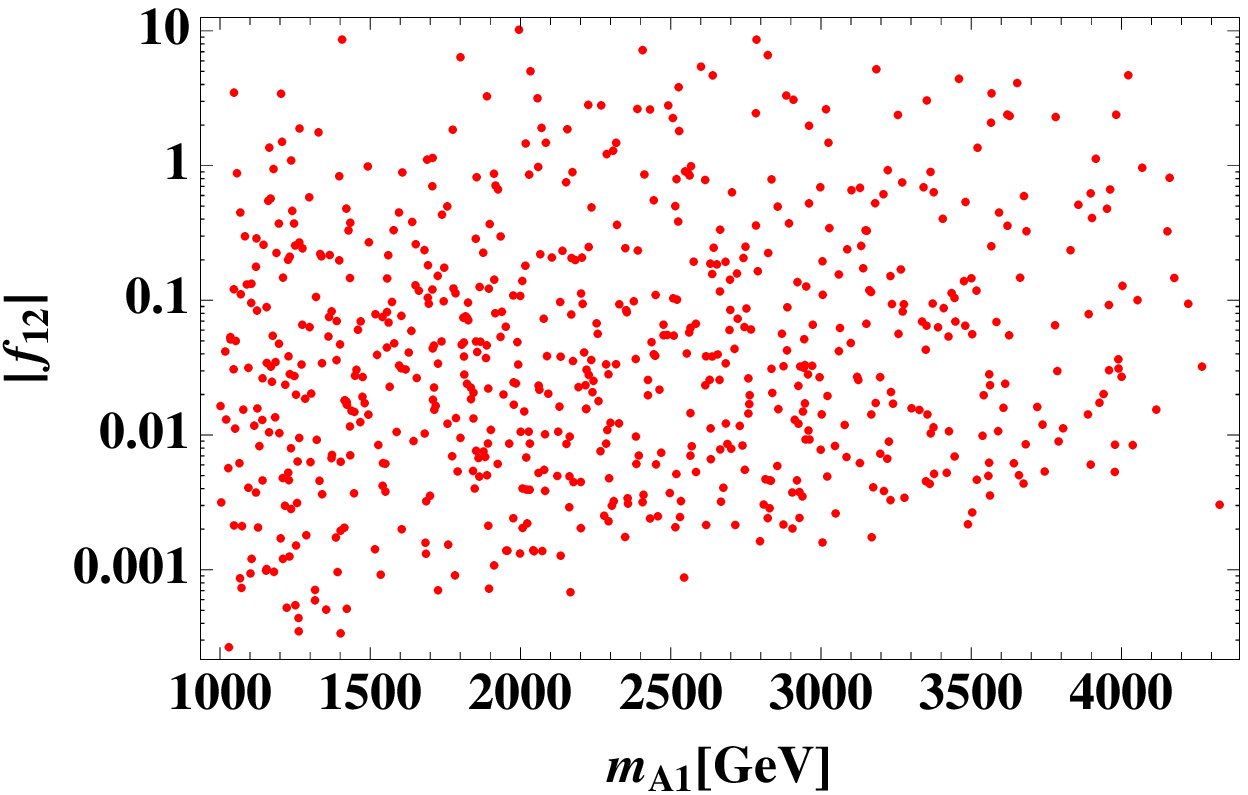}
\includegraphics[width=80mm]{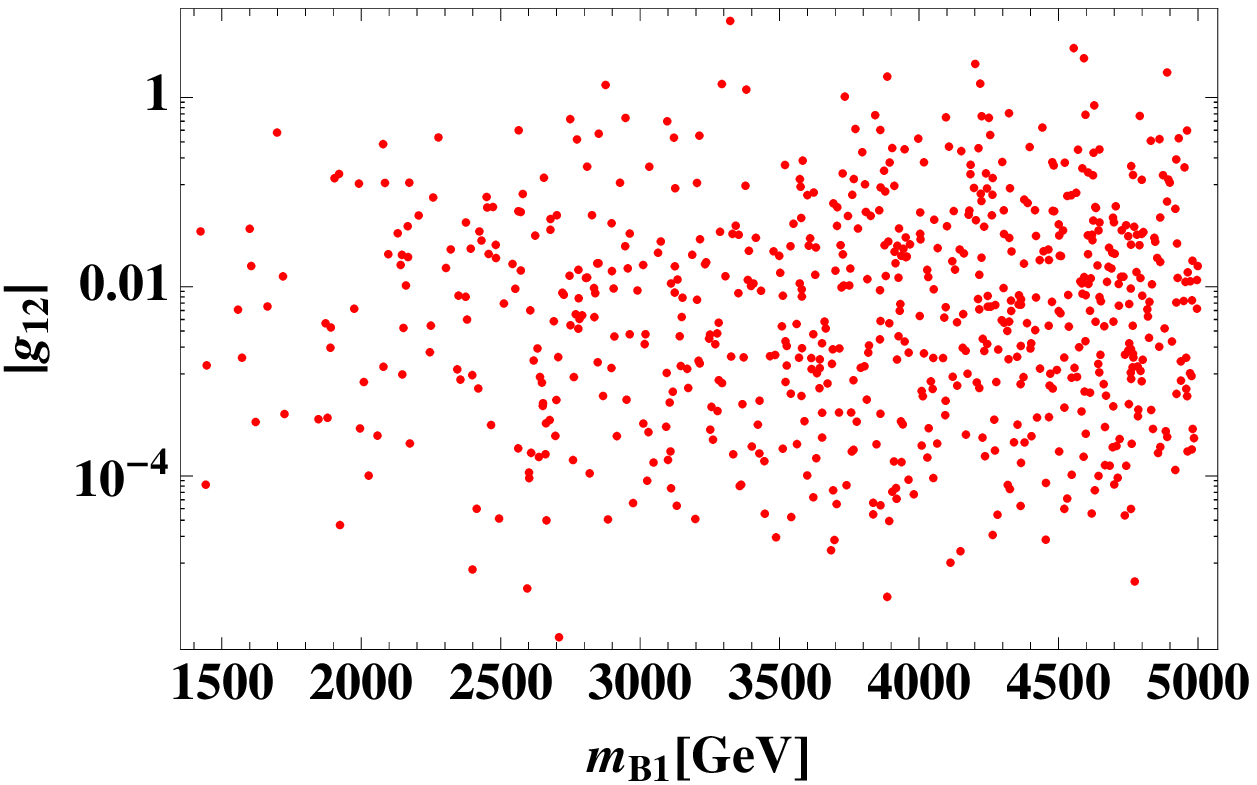}
\includegraphics[width=80mm]{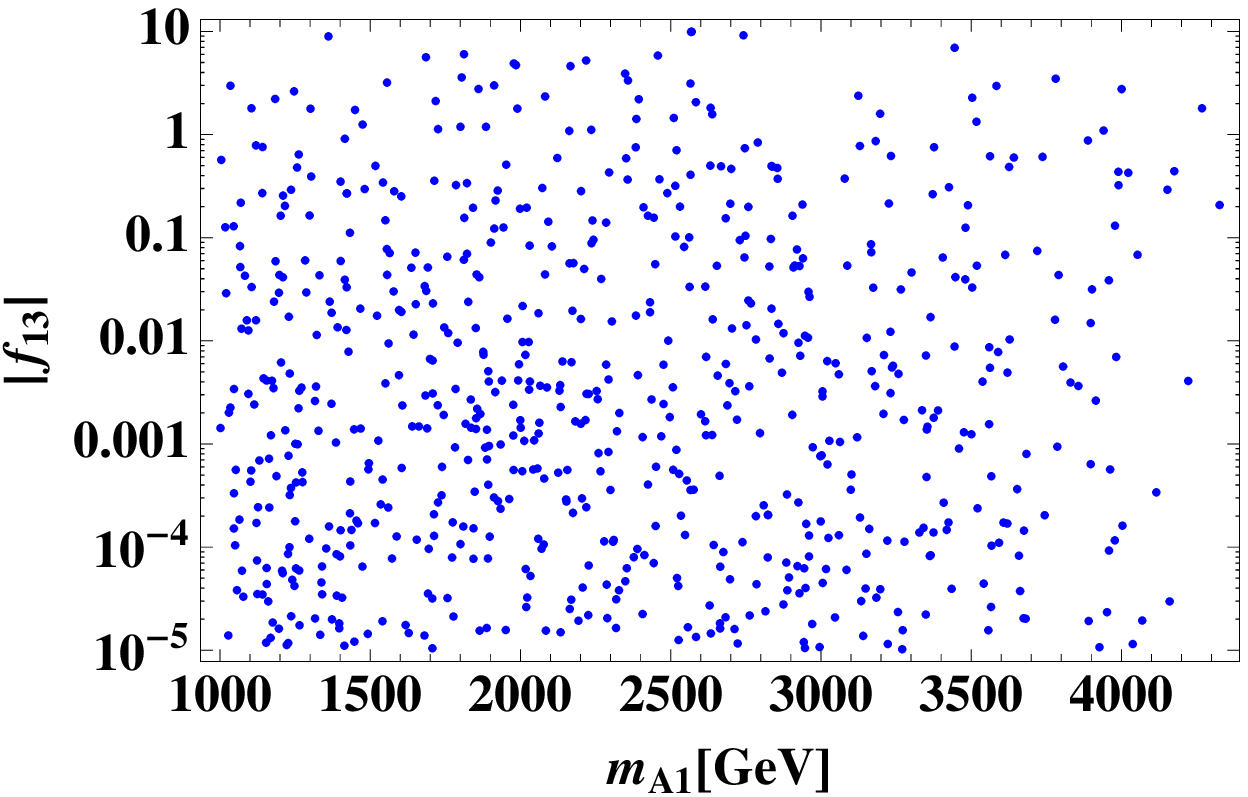}
\includegraphics[width=80mm]{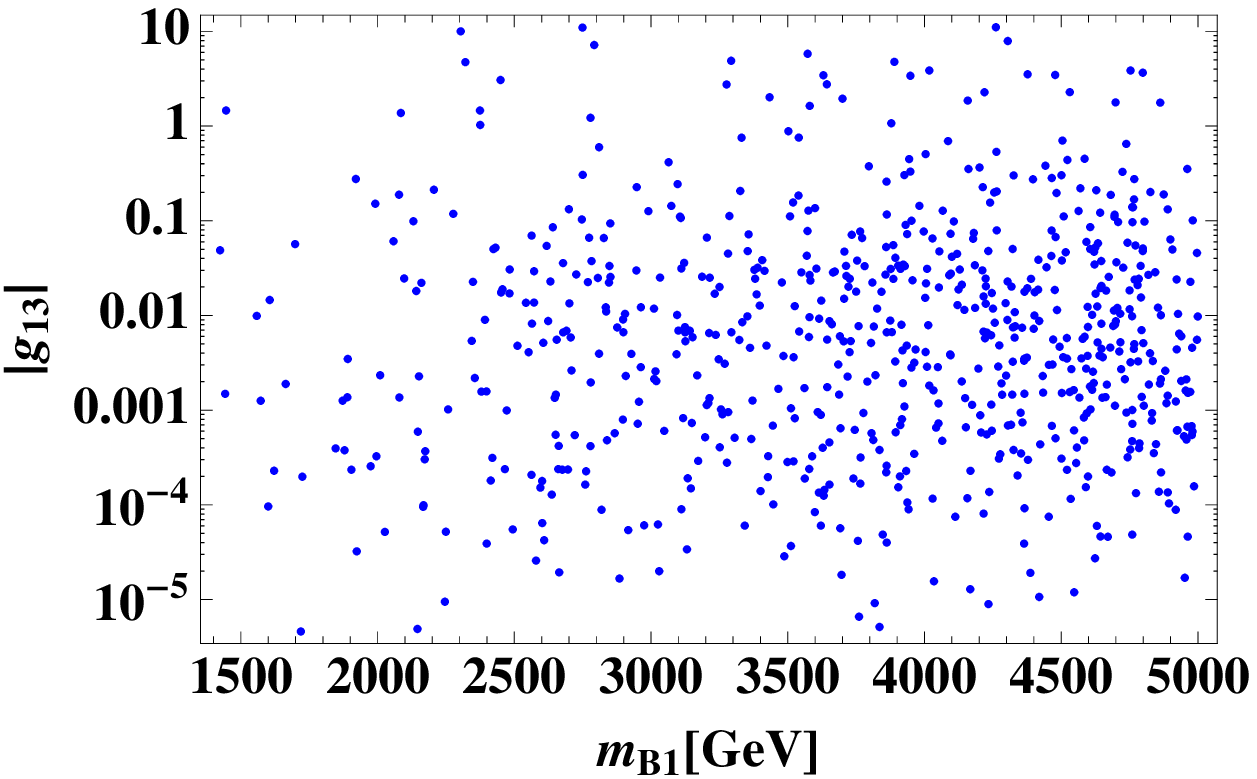}
\caption{Scatter plots of $f_{1j}$ versus $m_{A_1}=m_{A_2} \approx m_\eta$ (left 
panels), and $g_{1j}$ versus $m_{B_1}=m_{B_2} \approx m_\delta$ (right panels).
{
The vertical line of 1 TeV is superimposed due to the 
direct search limit of leptoquarks.}
}
\label{fg}
\end{figure}

\begin{figure}[t!]
\includegraphics[width=80mm]{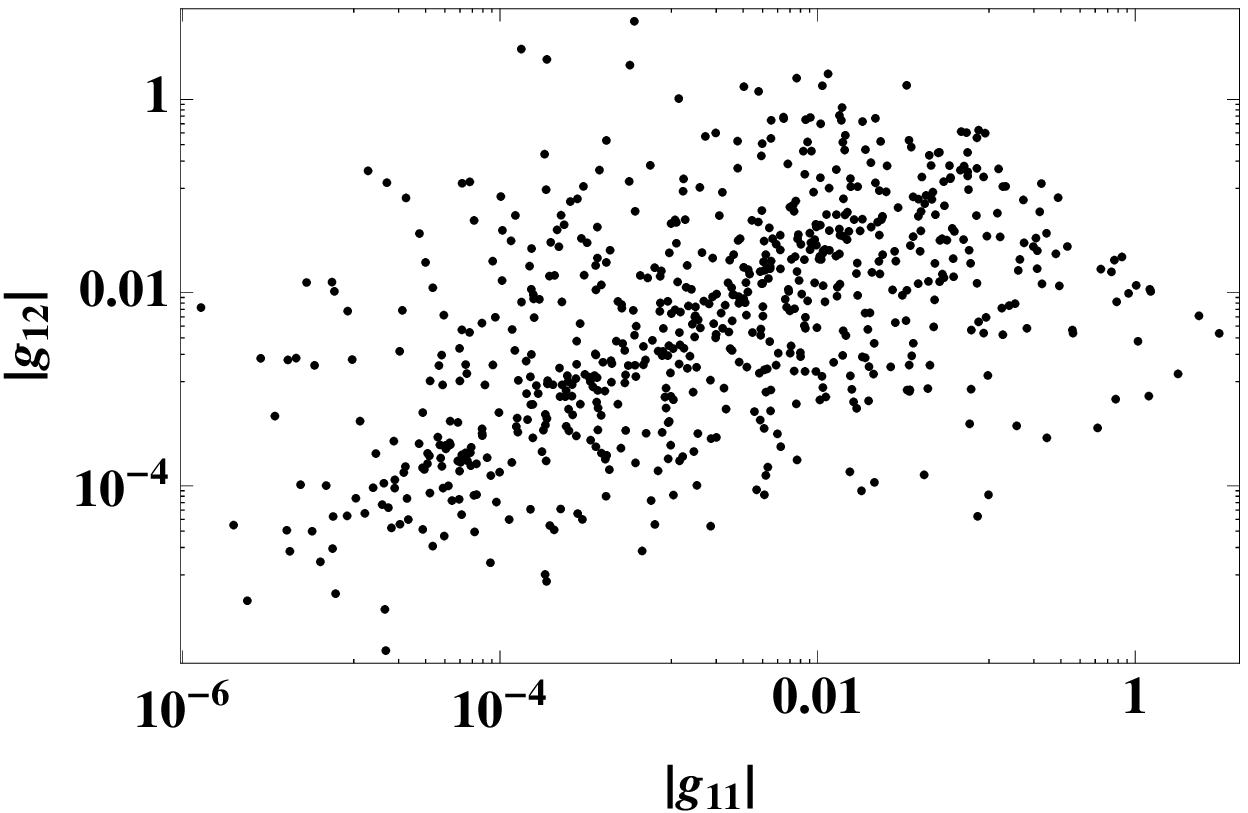}
\includegraphics[width=80mm]{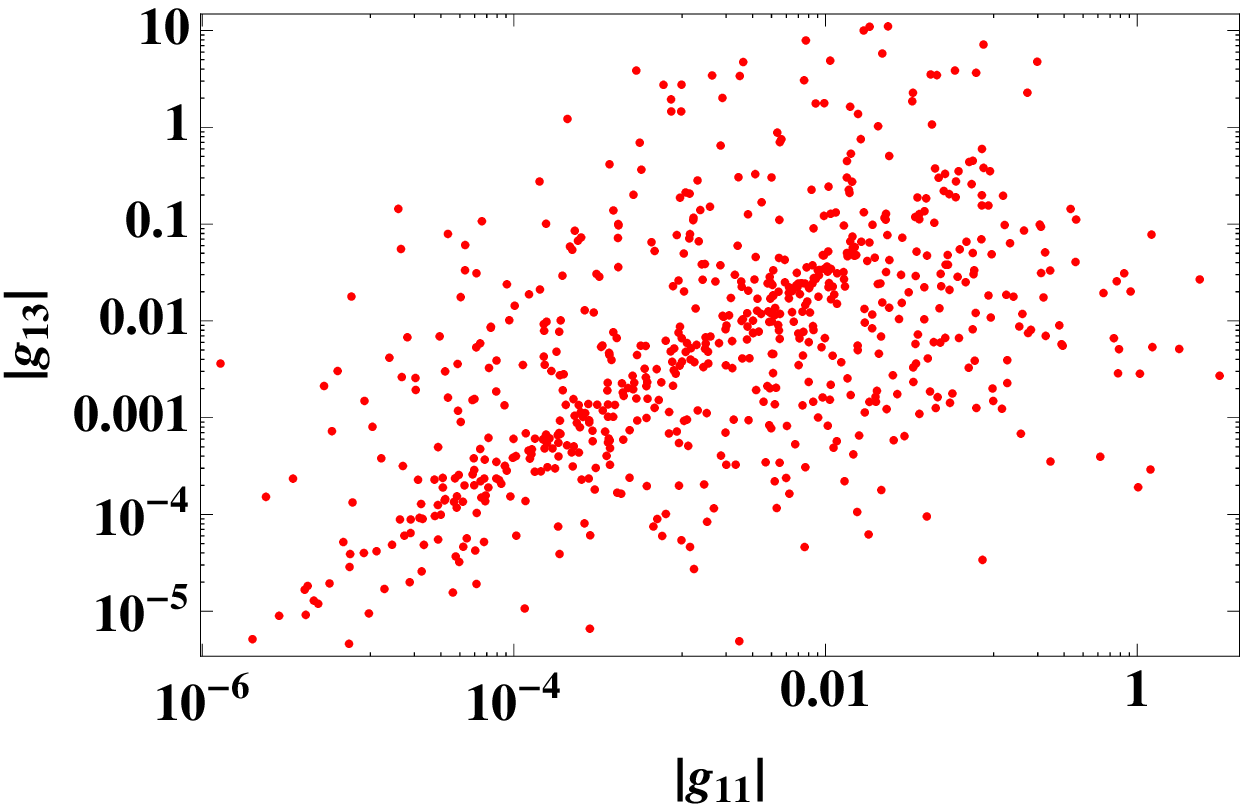}
\includegraphics[width=80mm]{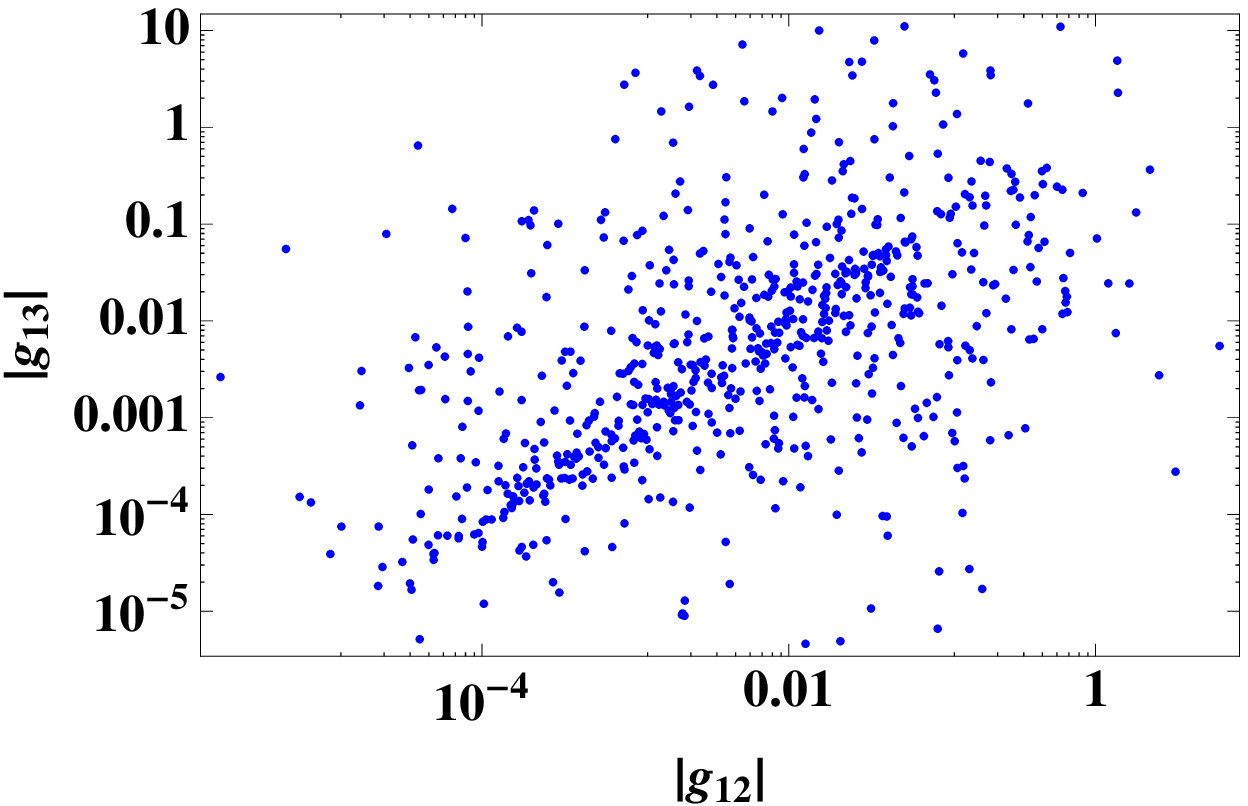}
\caption{{Scatter plots of $g_{11}$ versus $g_{12}$ with black dots
(upper left), $g_{11}$ versus $g_{13}$ with red dots (upper right), and 
$g_{12}$ versus $g_{13}$ with blue dots (lower). These trends mainly come 
from $\ell_a\to\ell_b\gamma$ at one-loop level. Especially, the upper left
panel (black dotted) is more restrictive than the other two panels, 
because both components $g_{12}$ and $g_{11}$ are related to the most 
stringent constraint of $\mu\to e\gamma$.
}}
\label{fg-gtrend}
\end{figure}

In the left panel of Fig.~\ref{fig:nums}, we show the allowed region in
$m_{A_1}$-$m_{B_1}$ plane, which suggests the relation
$m_{A_1}\lesssim m_{B_1}$ that is mainly required from the constraints
of $\Delta S$ and $\Delta T$.
We also put a vertical line of $m_{A_1} = 1$ TeV onto the
figure, because the collider limit on leptoquarks is roughly 1 TeV.
In the right panel of Fig.~\ref{fig:nums}, we show the allowed region
in $C_{9}$-$C_{9}'$ plane.  It suggests that the relation 
{$C_9(\in[0,0.2]) \ll C_9'(\in[0,0.7])$}
is realized for most of the parameter region while
{a parameter set provides} $C_9 \sim C_9'$; the cases of $C_9(={-}C_{10}) < 0$ and
$C_9'(=-C_{10}') < 0$ are disfavored by the constraints from $B_s$ and
$B_d$ decay branching ration Eqs.~(\ref{eq:Bsmumu}) and
(\ref{eq:Bdmumu}).
$C_9(={-}C_{10}) \sim C_9'(=-C_{10}')$ case is not favored by the global analysis.
In Fig.~\ref{fg}, we show the scatter plots of $f_{11,12,13}$ versus
$m_{A_1} = m_{A_2} \approx m_\eta$ (left panels), and
$g_{11,12,13}$ versus $m_{B_1} = m_{B_2} \approx m_\delta$ (right panels).
{$f_{11}$ and $f_{13}$ go over all the range, while $f_{12}$ is favor of rather larger value.} 
On the other hand, {$[g_{11},g_{12}]  \alt {\cal O}(1)$, while $g_{13}$ is likely to a free parameter}, as shown in Fig.~\ref{fg-gtrend}.

We note that the $\mu \to e \gamma$ process provides the strongest constraint which bounds the combinations of the couplings $|f_{2i}^\dagger f_{i1}|$ and $|g_{2i}^\dagger g_{i1}|$. Therefore, we can see
that the collider limits obtained from the Drell-Yan process in 
Eqs.~(\ref{eta-limit}) and (\ref{delta-limit})
are weaker than the direct search mass limits of leptoquarks ($\sim 1$ TeV).

\subsection{Collider Predictions}

The most striking signature of the model is the lepton-flavor 
violating production via the $\eta$ or $\delta$ bosons in the $t$-channel,
resulting in the final states of $e^\pm \mu^\mp$, $e^\pm \tau^\mp$, or
$\mu^\pm \tau^\mp$. The SM irreducible backgrounds to these final states are
negligible.  
Since the $\delta$ boson is in general heavier than the
$\eta$ boson, we use the subprocess $d \bar d \to \ell_i \ell_j$ 
via the exchange of the $\eta$ boson to estimate the event rates. 
We give the signal event rates in Table~\ref{e-mu-tau}, using 
the parameters $f_{11}=10^{-2}, f_{12}=f_{13} = 10^{-1}$  and 
$m_\eta = 1$ TeV at the 13 TeV LHC with 300 fb$^{-1}$ luminosity.
Naively, the production cross section for $\ell_i \ell_j$ is 
proportional to $| f_{1i} f_{1j} |^2$. If we choose $f_{11} = 10^{-1}$ instead
of $10^{-2}$ the event rate will increase by 100 times, although the
number of parameter points for $f_{11} =10^{-1}$ is considerably less.
Therefore, the event rates may be large enough for observation if the
Yukawa couplings $f_{1j}$ are of order $O(10^{-1})$.

As we have mentioned above, pair production cross sections for leptoquarks
have been calculated with NLO accuracy \cite{pair} and the
cross sections at 13 TeV LHC for 1 TeV $\eta$ or $\delta$ bosons is 
of order $O(10)$ fb. The final state consists of two leptons and two jets,
among which the corresponding lepton and jet will form an invariant-mass
peak.
On the other hand, the $\eta$ or $\delta$ bosons can also be singly 
produced with the subprocess $g q \to \eta \ell$ \cite{singly}, 
followed by the decay of $\eta$ into
a lepton and a quark. The amplitude for the production involves
a strong coupling and a Yukawa coupling ($f_{ij}$ for $\eta$ but $g_{ij}$
for $\delta$). Nevertheless, since the sizes of $f_{ij}$ and $g_{ij}$ are
very small because of the small neutrino mass, the production cross
section for single $\eta$ or $\delta$ is very suppressed.  We shall
not further consider this production mechanism.

\begin{table}[t!]
\caption{\small \label{e-mu-tau}
Event rates for the $e^\pm \mu^\mp$, $e^\pm \tau^\mp$, or 
$\mu^\pm \tau^\mp$ final states with exchange of the $\eta$ boson in
the subprocess $d\bar d \to e^- e^+$ 
at the 13 TeV LHC with 300 fb$^{-1}$ luminosity.
}
\medskip
\begin{ruledtabular}
\begin{tabular}{lccc}
 Inputs & $e^\pm \mu^\mp$ &  $e^\pm \tau^\mp$ & $\mu^\pm \tau^\mp$ \\
 $f_{11}=10^{-2}$, $f_{12}=f_{13}=10^{-1}$, $m_{\eta}=1$ TeV 
 & $0.057$  & $5.7$ & $0.057$
\end{tabular}
\end{ruledtabular}
\end{table}

\section{Conclusions}
We have proposed a simple extension of the SM with two leptoquark-like
scalar bosons $\eta$ and $\Delta$, which couple to all three generations
of fermions.  It can explain the neutrino masses and oscillations data,
and the most importantly explain the anomalies observed  in
$b \to s \ell^+ \ell^-$ including the lepton-universality violation and
angular distributions, and is at the same time consistent with all the
LFVs, FCNCs, Drell-Yan production, and collider searches. 

We offer a few more comments as follows.
\begin{enumerate}
\item The contributions of $\eta$ and $\Delta$ to the muon $g-2$ are
negligible compared to the experimental uncertainties.

\item The contributions of the $\eta_{2/3}$ to the Drell-Yan process,
proportional to $|f_{11}|^4$ ($|f_{12}|^4$) for $e^+ e^- \; (\mu^+ \mu^-)$
final state, will show up as an enhancement in the large invariant-mass
region.

\item The most interesting collider signature for the $\eta$ or $\delta$
boson is the lepton-flavor violating production such as 
$e^\pm \mu^\mp$, $e^\pm \tau^\mp$, and $\mu^\pm \tau^\mp$, which are
proportional to $|f_{11} f_{12}|^2$, $|f_{11} f_{13}|^2$, and 
$|f_{12} f_{13}|^2$, respectively. The event rates 
may be large enough for observation if the
Yukawa couplings $f_{1j}$ are of order $O(10^{-1})$.

\item The direct search limits on $\eta$ or $\delta$ bosons, just like
leptoquarks, are currently stronger than the indirect bounds from the
Drell-Yan process that we obtained in Eqs.~(\ref{eta-limit}) and 
(\ref{delta-limit}).

\end{enumerate}

\section*{Acknowledgments}
This work was supported by the Ministry of Science and Technology
of Taiwan under Grants No. MOST-105-2112-M-007-028-MY3.

\begin{appendix}

\section{New particle contributions to vacuum polarization diagram}
Here we summarize the contributions to $\Pi_{\pm}(q^2)$, $\Pi_{33}(q^2)$, $\Pi_{3Q}(q^2)$ and  $\Pi_{QQ}(q^2)$ in Eq.~(\ref{eq:piZ}) and (\ref{eq:piW}) from new particles in our model.

\noindent
{\bf Contributions to $\Pi_{\pm}(q^2)$ } \\
The one loop contributions from three-point gauge interaction are denoted by $\Pi_\pm^{XY}(q^2)$ where $X$ and $Y$ indicate the particles inside the loop.
They are summarized as follows;
\begin{align}
& \Pi_{\pm}^{A_1(B_1) \delta_{4/3}} (q^2) = \frac{2}{(4\pi)^2} s_{\alpha_1}^2(c_{\alpha_1}^2) G(q^2, m_{A_1(B_1)}^2, m_\delta^2), \\
& \Pi_{\pm}^{B_1 \delta_{4/3}} (q^2) = \frac{2}{(4\pi)^2} c_{\alpha_1}^2 G(q^2, m_{B_1}^2, m_\delta^2), \\
& \Pi_{\pm}^{A_1A_2} (q^2) = \frac{2}{(4\pi)^2} \left( s_{\alpha_1} s_{\alpha_2} - \frac{1}{\sqrt{2}} c_{\alpha_1} c_{\alpha_2} \right)^2 G(q^2, m_{A_1}^2, m_{A_2}^2), \\
& \Pi_{\pm}^{B_1B_2} (q^2) = \frac{2}{(4\pi)^2} \left( c_{\alpha_1} c_{\alpha_2} - \frac{1}{\sqrt{2}} s_{\alpha_1} s_{\alpha_2} \right)^2 G(q^2, m_{B_1}^2, m_{B_2}^2),  \\
& \Pi_{\pm}^{A_1B_2} (q^2) = \frac{2}{(4\pi)^2} \left( s_{\alpha_1} c_{\alpha_2} + \frac{1}{\sqrt{2}} c_{\alpha_1} s_{\alpha_2} \right)^2 G(q^2, m_{A_1}^2, m_{B_2}^2), \\
& \Pi_{\pm}^{B_1A_2} (q^2) = \frac{2}{(4\pi)^2} \left( c_{\alpha_1} s_{\alpha_2} - \frac{1}{\sqrt{2}} s_{\alpha_1} c_{\alpha_2} \right)^2 G(q^2, m_{B_1}^2, m_{A_2}^2), 
\end{align}
where 
\begin{align}
&  G(q^2, m_P^2, m_Q^2) =  \int dx dy \delta (1-x-y) \Delta_{PQ} [\Upsilon+1 - \ln \Delta_{PQ}], \nonumber \\
 & \Delta_{PQ} = -q^2 x(1-x) + x m_{P}^2 + y m_{Q}^2, \quad \Upsilon = \frac{2}{\epsilon} - \gamma - \ln (4 \pi).
\end{align}
The one loop contributions from four-point gauge interaction are denoted by $\Pi_\pm^{X}(q^2)$ where $X$ indicates the particle inside the loop.
They are summarized as follows;
\begin{align}
& \Pi_\pm^{A_{1}}(q^2) = - \frac{1}{2 (4\pi)^2} (4 s_{\alpha_{1}}^2+ c_{\alpha_{1}}^2) H(m_{A_{1}}^2), \\
& \Pi_\pm^{A_{2}}(q^2) = - \frac{1}{2 (4\pi)^2} (2 s_{\alpha_{2}}^2+ c_{\alpha_{2}}^2) H(m_{A_{2}}^2), \\
& \Pi_\pm^{B_{1}}(q^2) = - \frac{1}{2 (4\pi)^2} (4 c_{\alpha_{1}}^2+ s_{\alpha_{1}}^2) H(m_{B_{1}}^2), \\
& \Pi_\pm^{B_{2}}(q^2) = - \frac{1}{2 (4\pi)^2} (2 c_{\alpha_{2}}^2+ s_{\alpha_{2}}^2) H(m_{B_{2}}^2), \\
& \Pi_\pm^{\delta_{4/3}}(q^2) = - \frac{1}{(4\pi)^2} H(m_{\delta}^2),
\end{align}
where
\begin{equation}
H(m_P^2) = m_P^2 [\Upsilon + 1 - \ln m_P^2].
\end{equation}

\noindent
{\bf Contributions to $\Pi_{33}(q^2)$, $\Pi_{3Q}(q^2)$ and $\Pi_{QQ}(q^2)$ } \\
The one loop contributions from three-point gauge interaction are denoted by $\Pi_{33,3Q,QQ}^{XY}(q^2)$ where $X$ and $Y$ indicate particles inside the loop.
They are summarized as follows;
\begin{align}
& \Pi_{[33,3Q,QQ]}^{A_1 A_1} = \frac{2}{(4\pi)^2} \left[ \frac{1}{4} c_{\alpha_1}^4, \frac{1}{6} c_{\alpha_1}^2, \frac{1}{9} \right] G(q^2, m_{A_1}^2, m_{A_1}^2), \\
& \Pi_{[33,3Q,QQ]}^{B_1 B_1} = \frac{2}{(4\pi)^2} \left[ \frac{1}{4} s_{\alpha_1}^4, \frac{1}{6} s_{\alpha_1}^2, \frac{1}{9} \right] G(q^2, m_{B_1}^2, m_{B_1}^2), \\
& \Pi_{[33,3Q,QQ]}^{A_1 B_1} = \frac{2}{(4\pi)^2} \left[ \frac{1}{4} s_{\alpha_1}^2 c_{\alpha_1}^2, 0, 0 \right] G(q^2, m_{A_1}^2, m_{B_1}^2)  \\
& \Pi_{[33,3Q,QQ]}^{A_2 A_2} = \frac{2}{(4\pi)^2} \left[  s_{\alpha_2}^4 - s_{\alpha_2}^2 c_{\alpha_2}^2 + \frac{1}{4} c_{\alpha_2}^4, \frac{2}{3} s_{\alpha_2}^4 - s_{\alpha_2}^2 c_{\alpha_2}^2 + \frac{1}{3} c_{\alpha_2}^4, \frac{4}{9} (s_{\alpha_2}^2 - c_{\alpha_2}^2)^2 \right] G(q^2, m_{A_2}^2, m_{A_2}^2), \\
& \Pi_{[33,3Q,QQ]}^{B_2 B_2} = \frac{2}{(4\pi)^2} \left[  c_{\alpha_2}^4 - s_{\alpha_2}^2 c_{\alpha_2}^2 + \frac{1}{4} s_{\alpha_2}^4, \frac{2}{3} c_{\alpha_2}^4 - s_{\alpha_2}^2 c_{\alpha_2}^2 + \frac{1}{3} s_{\alpha_2}^4, \frac{4}{9} (s_{\alpha_2}^2 - c_{\alpha_2}^2)^2 \right] G(q^2, m_{B_2}^2, m_{B_2}^2), \\
& \Pi_{[33,3Q,QQ]}^{A_2 B_2} = \frac{2}{(4\pi)^2}  s_{\alpha_2}^2 c_{\alpha_2}^2 \left[  \frac{9}{4} , 3, 4 \right] G(q^2, m_{A_2}^2, m_{B_2}^2), \\
& \Pi_{[33,3Q,QQ]}^{\delta_{4/3} \delta_{4/3}} = \frac{2}{(4\pi)^2} \left[ 1, \frac{4}{3}, \frac{16}{9} \right] G(q^2, m_{\delta}^2, m_{\delta}^2). 
\end{align}
The one loop contributions from four-point gauge interaction are denoted by $\Pi_{33,3Q,QQ}^{X}(q^2)$ where $X$ indicates the particle inside the loop.
They are summarized as follows;
\begin{align}
& \Pi_{[33,3Q,QQ]}^{A_1} = - \frac{2}{(4\pi)^2} \left[s_{\alpha_1}^2 + \frac{1}{4} c_{\alpha_1}^2, \frac{2}{3} s_{\alpha_1}^2+\frac{1}{6} c_{\alpha_1}^2, \frac{4}{9} s_{\alpha_1}^2 + \frac{1}{9} c_{\alpha_1}^2 \right] H( m_{A_1}^2), \\
& \Pi_{[33,3Q,QQ]}^{B_1} = - \frac{2}{(4\pi)^2} \left[c_{\alpha_1}^2 + \frac{1}{4} s_{\alpha_1}^2, \frac{2}{3} c_{\alpha_1}^2+\frac{1}{6} s_{\alpha_1}^2, \frac{4}{9} c_{\alpha_1}^2 + \frac{1}{9} s_{\alpha_1}^2 \right] H( m_{B_1}^2), \\
& \Pi_{[33,3Q,QQ]}^{A_2} = - \frac{2}{(4\pi)^2} \left[ \frac{1}{4} c_{\alpha_2}^2, \frac{1}{3} c_{\alpha_2}^2, \frac{1}{9} s_{\alpha_2}^2 + \frac{4}{9} c_{\alpha_2}^2 \right] H( m_{A_2}^2), \\
& \Pi_{[33,3Q,QQ]}^{B_2} = - \frac{2}{(4\pi)^2} \left[ \frac{1}{4} s_{\alpha_2}^2, \frac{1}{3} s_{\alpha_2}^2, \frac{1}{9} c_{\alpha_2}^2 + \frac{4}{9} s_{\alpha_2}^2 \right] H( m_{B_2}^2), \\
& \Pi_{[33,3Q,QQ]}^{\delta_{4/3}} = - \frac{2}{(4\pi)^2} \left[ 1, \frac{4}{3} , \frac{16}{9} \right] H( m_{\delta}^2). 
\end{align}

\end{appendix}


\end{document}